# Pyroelectric Control of Spin Polarization Assisted Coexistence of Giant Positive and Negative Magnetocaloric Effects

Gaurav Vats[a,b,c]*, Ashok Kumar[d#], Ram S. Katiyar [e#] and Nora Ortega[e#]

[a]*IITB-Monash Research Academy, Indian Institute of Technology Bombay, Powai, India 400076*

[b]*Department of Mechanical Engineering, Indian Institute of Technology Bombay, Powai, India 400076*

[c]*Department of Mechanical and Aerospace Engineering, Monash University, Clayton, VIC 3800, Australia*

[d]*CSIR-National Physical Laboratory, Dr. K. S. Krishnan Marg, Delhi, India 110012*

[e]*Department of Physics and Institute for Functional Nanomaterials, University of Puerto Rico, San Juan, PR 00931-3343, USA*

**\*Email:** gvats17@iitb.ac.in **Phone: +91-22-2572 2545 Ext. 7500, Fax: +91-22-2572 8675**

[#]Authors contributed equally.

**Abstract:**

Electric field control of magnetism is the key to many next generation spintronics applications[1]. Ferroelectric control of spin polarization[2,3] followed by the electrically driven repeatable magnetization reversal in the absence of applied magnetic field[4] have proven to be the milestones in this direction. This article propose how these phenomena could be utilized in a reverse manner to gain control over magnetization even in the absence of electric field. In turn, coexistence of positive[5] as well as negative[6] magnetocaloric effect (MCE) is attained in tri-layered $PbZr_{0.53}Ti_{0.47}O_3$-$CoFe_2O_4$-$PbZr_{0.53}Ti_{0.47}O_3$ (PZT/CFO/PZT) nanostructures for identical temperature ranges when subjected to different applied magnetic fields. Unlike conventional approaches[5-9] the present study demonstrate that it is possible to obtain giant MCE merely by magneto-pyroelectric coupling. Consequently, the MCE entropy changes calculated using Maxwell equations are found to be as large as reported for existing giant MCE values[6,9,10].

Coherent magnetization[11] or electric-field-assisted[12] switching have expanded the horizon of magneto-electric coupling (MEC) for low-power electrical control of modern spintronics devices[1]. Literature consists of several such trials as electrical control of magnetic anisotropy[13], domain structure[4,14], spin polarization[2,15] or phase transitions[16] in thin ferromagnetic metal films[17], ferromagnetic semiconductors[18-20], multiferroics[21,22] and their hybrids[23]. Multiferroic heterostructures made by conventional ferromagnets and ferroelectrics[24,25] stands out, especially, and are considered among the most promising candidates in this domain. Research on these gained renewed interest after Tsymbal *et. al.* demonstrated the ferroelectric control of magnetism in Fe/BaTiO$_3$ (BTO)[3] and tunneling across ferroelectric interface[26]. Later, theoretical studies on, Co$_2$MnSi[27] and Fe$_3$O$_4$[28] also illustrated large changes in magnetic moments at ferroelectric-ferromagnet interfaces. These predictions attracted interests for novel spintronics devices such as magnetic random access memories and spin field-effect transistors[29] which became an inspiration to develop the solution for a non-volatile control of spin polarization. This was experimentally achieved by reversing the electrical polarization of ultrathin ferroelectric BTO in Fe/BTO/ La$_{0.67}$Sr$_{0.33}$MnO$_3$ (LSMO) tunnelling junctions[2]. Currently, exchange-bias coupling[30-32], interface bonding driven MEC[3], spin-dependent screening[33] and strain-mediated coupling[34-36] are among the popularly explored mechanisms for having such control. Recently, we have also contributed to develop an understanding of the temperature dependence of MEC[37,38]. In addition, a non-volatile repeatable magnetization reversal with no applied field has also been demonstrated by Ghidini, M. *et al.*[4]. This paper utilize the same mechanism but in a reverse manner. It seeks to manifest the control over magnetization even in the absence of an electric field using PbZr$_{0.53}$Ti$_{0.47}$O$_3$-CoFe$_2$O$_4$-PbZr$_{0.53}$Ti$_{0.47}$O$_3$ (PZT/CFO/PZT) tri-layered nanostructures. This causes huge fluctuations in magnetization *(M)* with a change in temperature and applied magnetic field *(H)* leading to coexistence of positive[5] as well as negative[6,39] magnetocaloric effect (MCE) in identical temperature ranges. These tri-layered nanostructures have been well explored for their active device applications[40-45] and a brief review of the experimental details are provided in the following paragraph[43-45].

The tri-layered nanostructures were deposited on $La_{0.5}Sr_{0.5}CoO_3$ (LSCO) coated (100) MgO substrate (see inset Figure 1) using pulsed laser deposition technique. The physical characterizations were carried out using X-ray diffraction and micro-Raman spectra (See supplementary Figures S1 and S2). The presence of distinct PZT and CFO peaks without the appearance of any additional peak is depicted which reveals that individual ferroelectric and ferromagnetic phases were retained in the configuration. Further, a high-resolution transition electron microscope (TEM) was used to investigate the formation of PZT/CFO/PZT structures and the inter-diffusion of constituent PZT and CFO layers (Figure S3). Thereafter, electrical measurements were performed and the temperature dependent *P-E* loops thus obtained are highlighted in Figure 1 (a). Figure 1 (b) shows the temperature dependence of the twice remanent polarization. Interestingly, it has been observed that these MLNs illustrate an unusual ferroelectric behavior. Surprisingly, the polarization falls with a decrease in temperature while generally the phenomenon is reversed. A nominal change in polarization with huge fluctuation in frequency eliminates the possibility of surface/interface effects and therefore this unusual shift has been attributed to dominance of dynamic MEC at lower temperatures[43,44]. Subsequently, the magnetic measurements were performed parallel to the c-axis (applied field was perpendicular to the substrate). Figure 2 shows the temperature dependant magnetization versus field (*M-H*) behavior of PZT/CFO/PZT in the first quadrant for selected temperatures while the complete loops are highlighted in the inset. The configuration is found to show almost negligible magnetic hysteresis. Additionally, analogous to electrical polarization the magnetization behavior has also been found peculiar for magnetic fields applied below 7 kOe. Beyond this the magnetization is found to follow the general trend of decrease in magnetization with increase in temperature (Note the change in magnetization with temperature below and above 7 kOe in Figure 2 and 3).

It is to be noted that the magnetization contribution of the LSCO buffer layer was merely 2–4% of the multilayer structure. Besides, the contribution is subtracted from the magnetization value of tri-layered nanostructures the temperature variation of the remanent ($\Delta M_r = M_r(PZT/CFO/PZT) - M_r(LSCO)$) and saturation ($\Delta M_s = M_s(PZT/CFO/PZT) - M_s(LSCO)$) magnetization indicates that $\Delta M_s$ consistently maintain a value of 0.4-0.5 MAm$^{-1}$ while the $\Delta M_r$ linearly falls from 60 to 10 kAm$^{-1}$ over the temperature spectrum of 150 K to 400 K[44]. $\Delta M_s$ of theses nanostructures are two to three times higher in contrast to CFO thin films[46,47] and CFO-based multilayered structures[48]. At the same time $\Delta M_r$ values are suppressed by at least a factor of half from the equivalent counterpart. Since the configuration is independent of any structural transitions and space charge effects the only plausibility for such behavior is dynamic MEC[44]. Ortega et. al. shed light on this using the Ginzburg–Landau theory for the second-order phase transition of ferroelectromagnets and showed that the MEC in PZT/CFO/PZT is stronger at lower temperatures and becomes weak at higher temperatures due to a substantial fall in remanent magnetization at elevated temperatures[44]. However, PZT is also a good pyroelectric material with a pyroelectric coefficient of 380 µCm$^{-2}$K$^{-1}$ [49-51]. This means that even the in absence of an applied electric field it can exhibit huge fluctuations in electrical polarization (see variation in remanent polarization (2$P_r$) with temperature in Figure 1 (b)). $P_r$ first increases with temperature till 200 K and starts dropping thereafter. Therefore, the effect of $P_r$ is negligible at temperatures below 150 K. It is also evident from the magnetization loops below 150 K (between 0 and 7 kOe) which are having magnetizations higher relative to that at 200 K irrespective of the applied magnetic field. These become an exception to the trend and are highlighted in magnified *M-H* isotherms of Figure 3 (a) for applied field below 5 kOe and (b) above 5 kOe (inset depicts the overall trend for all temperatures; star symbol indicates *M-H* isotherms that are independent of MEC and become an exception in the respective applied field range). This clearly interprets that the trend of magnetization below 7 kOe is dominated by pyroelectric effect for the temperatures above 150 K. The degree of this dominance reaches its maximum limit at 200 K; thereafter it starts decreasing. Another key observation is that magnetic fields applied beyond 7 kOe are sufficient to overcome the effect of electrical polarization and consequently the magnetization falls with the increase in temperature (Figure 3 (b)). This is due to dominance of the cumulative magnetization of CFO and LSCO above 7 kOe. Since the magnetization measurements were performed in the absence of applied electric field it would be prudent to label the effect as 'pyroelectric control of magnetization' rather than MEC. Intriguingly, the effect is found to be completely recoverable with excellent fatigue properties tested over $10^8$ cycles[44]. The aforementioned phenomena with colossal magnetization, high breakdown strength and low losses make the PZT/CFO/PZT ideal for MCE investigations.

Magnetocaloric effect is the isothermal entropy/heat change procured by adiabatic tuning of magnetization[8]. The effect was primarily exploited for laboratory applications[52] until giant MCE was discovered using reversible distortions in symmetry[5-9,53,54] or volume[55] aided by strongly coupled magnetic and structural transitions[10]. Table 1 reviews the existing giant MCE in selected materials. In general the cooling is obtained during demagnetization and termed as positive MCE[56]. However, if the cooling is obtained while magnetizing the material the effect is said to be inverse or negative MCE[56,57]. A more detailed information about the physics beyond negative caloric effects can be found in[39]. Both effects can be distinguished using negative (positive MCE) and positive values (inverse or negative MCE) of entropy change *(ΔS(H))* obtained from the Maxwell relations. The relation suggests that the MCE entropy change *(ΔS(H))* for an applied magnetic field *(H)* can be indirectly estimated as[10,58]

$$\Delta S(H) = \mu_0 \int_0^H \left(\frac{\partial M}{\partial T}\right)_{H'} dH' \qquad (1)$$

Here $\mu_0$ is the permeability of the free space. The $\left(\frac{\partial M}{\partial T}\right)_{H'}$ can be obtained through a set of *M (H)* isotherms obtained from thermodynamic equilibrium. Further, the corresponding MCE temperature change (Δ*T*) can be determined by[10,58]

$$\Delta T = -\frac{T}{C}\Delta S(H) = -\frac{T\mu_0}{C}\int_0^H \left(\frac{\partial M}{\partial T}\right)_{H'} dH' \qquad (2)$$

*C* is the specific heat capacity of the material and can be assumed constant in the absence of transitions. This technique is well recognized and independent of microscopic details[10]. We have obtained *M(T)* values corresponding to *H* varying from 0 to 0.5 T (Figure 3 (a)) and 0.5 T to 2 T (Figure 3 (b)) using standard procedures from the upper branches of the *M-H* curve of the first quadrant (*H* >0). These values are used to calculate the Δ*S(H)* and Δ*T* using equations 1 and 2 respectively. Figure 4 (a) and (b) shows the variation in entropy *(ΔS(H))* as a function of temperature for fields applied below and above 0.5 T. Figure 4 (c) and (d) disclose the MCE (indirect estimation using Maxwell relations) temperature changes *(ΔT)* for the respective ranges of *H*. Since these nanostructures show large change in magnetization in the temperature range of 150 K to 400 K, the MCE investigation are restricted within this temperature range. Also, the effect of pyroelectric control of spin can be noticed in this domain. The MCE calculated using indirect mode of measurement is found to be negative for *H*<0.5 T. The entropy first increases with rise in applied field till 0.4 T ($\Delta S(H)_{max}$=1.5 Jkg$^{-1}$K$^{-1}$ at T=245 K) and thereafter a drop in $\Delta S(H)$ is noticed for 0.5 T (Figure 4 (a)). This regime corresponds to the dominance of MEC by pyroelectric effect. From the existence of MCE maxima at 0.4 T it can be interpreted that the pyroelectric control of spin is maximum corresponding to 0.4 T. Beyond this the pyroelectric governance starts falling and completely vanishes at 0.7 T. This is again reflected by the $\Delta S(H)$ plots of MCE for applied fields above 0.5 T. Almost all $\Delta S(H)$ values are shifted in the positive MCE regime and the reversal of this shift can beautifully observed in Figure 4 (b). The $\Delta S(H)$ maxima for $H$=0.75 T is slightly positive and convex in shape. Though the next curve for 1 T is entirely negative, it has maintained its convexity. Subsequently, the $\Delta S(H)$ is entirely independent of pyroelectric control of spin and curves become concave for all applied fields above 1 T. In general, from the perspective of practical applications it is suggested to explore the MCE for fields below 2 T. Hence the $\Delta S(H)$ for $H$>2 T are not shown here. But, the trend is found to be consistent (concave shape is maintained) even for higher applied fields. The positive MCE limit for the $H$=2 T is found to be 18.15 Jkg$^{-1}$K$^{-1}$ corresponding to 245 K. These values are significantly higher than many existing giant values[6,10] (see Table 1 for comparison). Figure 4 (c) and (d) augments the concurrent $\Delta T$ (using equation 2) as a function of temperature and applied field. The trend in $\Delta T$ versus $T$ plots are analogous to that of obtained for $\Delta S(H)$. The maximum $\Delta T$ are rendered to be 14.9 K (positive MCE at T=300 K) and 1.17 K (negative MCE at T=266 K). However, it is to be noted that these values are not exact and depend upon equilibrium thermodynamics as polarization decays with time[59]. This decay could be of 50 percent of the polarization values measured in short time scale[59]. Coexistence of positive as well as negative MCE in the identical temperature ranges makes for better device prospects if operated in a cyclic fashion using conventional refrigeration cycles[60-65] or magnetic version of reversed Olsen cycle[66]. Moreover, the configuration is independent of structural transitions[5-9,53-55] which are often responsible for giant MCE. Therefore, in future the nanostructures understudy can be tuned to obtain pyroelectric control over magnetic spin in the vicinity of the structural transitions so as to recover an enhanced MCE. Also, it will be advantageous to operate such systems with coupled electro-magnetocaloric effects. We are working in this direction and hope to come up with a solution soon. It is expected that the present work will help open up new opportunities in the area of solid state devices for the discovery of giant MCE as well as other suitable applications of pyroelectric control over magnetic spin and MEC.

**Experimental Details**

**Physical Characterization**

The tri-layered nanostructures of PZT/CFO/PZT were deposited from individual PZT and CFO targets on $La_{0.5}Sr_{0.5}CoO_3$ (LSCO) coated (100) MgO substrate using pulsed laser deposition technique. These ceramic targets of 2 cm diameter were initially prepared by the conventional solid-state route and thereafter an excimer laser (KrF, 248nm) with a laser energy density of 1.5 $Jcm^{-2}$ and pulse repetition rate of 10 Hz were used for deposition. The deposition process was carried out at an oxygen pressure of 13.33 Pa and the substrate temperature was maintained at 400°C. Afterwards the deposited films were annealed at 700 °C for 150 s using a rapid thermal annealing furnace. The total thickness of deposited films was found to be ~350nm. Further characterization were performed using by X-ray diffraction (XRD) technique with Cu Kα radiation in a Siemens D500 spinel CFO (overlapping with MgO peaks), (00l), (002) from the LSCO bottom electrode and (002) from oriented substrate. A more detailed description of characterization can be found elsewhere[44].

**Electrical and Magnetic Measurements**

The electrical measurements were carried out using Pt sputtered top electrodes with a diameter of ~200μm through a shadow mask and temperature dependent impedance measurements were carried out at an interval of 20 K in the range from 150 to 400 K. Subsequently, the temperature dependence of magnetic behavior is studied using a Quantum Design MPMS XL-7 SQUID magnetometer. The sample was placed in a standard drinking straw sample holder and the sample space was evacuated multiple times to ensure removal of air by displacing it with He. Magnetic measurements were carried out parallel to the c-axis (applied field was perpendicular to the substrate).

**Acknowledgment**

Authors would like to thank Professor J F Scott for help in editing the text.

**Table 1:** Comparison of MCE in selected compositions.

| Material | T (K) | $\Delta S(H)/\mu_0 H$ ($Jkg^{-1}K^{-1}T^{-1}$) |
|---|---|---|
| $Gd_5Si_1Ge_3$[54] | 136 | -13.6 |
| $MnCoGeB_{0.02}$[67] | 287 | -9.5 |
| LCMO‖BTO[10] | 190 | -9 |
| MnAs[68] | 318 | -6.4 |
| LaFe$_{11.7}$Si$_{1.3}$[55] | 184 | -6 |
| LaFe$_{11.57}$Si$_{1.43}$H$_{1.3}$[55] | 291 | -5.6 |
| Gd$_5$Si$_2$Ge$_2$[8] | 276 | -3.7 |
| Ni$_{52.6}$Mn$_{23.1}$Ga$_{24.3}$[53] | 300 | -3.6 |
| *Ni$_{50}$Mn$_{37}$Sn$_{13}$[6] | 299 | 3.8 |
| *Ni$_{50}$Mn$_{34}$In$_{16}$[57] | 190 | 2.4 |
| *CoMnSi$_{0.95}$Ge$_{0.05}$[56] | 215 | 1.8 |
| *PZT/CFO/PZT (Present Study) | 245 | 3.04 |
| PZT/CFO/PZT (Present Study) | 245 | -7.28 |

*Inverse MCE. The inverse effect obtained for PZT/CFO/PZT is corresponding to $H = 0.4$ T while the positive MCE is for $H = 2$ T.

**Figure Captions**

Figure 1: (a) Temperature dependant polarization versus electric field (*P-E*) and (b) twice remanent polarization behavior of PZT/CFO/PZT tri-layered nanostructure (Inset shows the deposited structure).

Figure 2: Temperature dependant magnetization versus applied field (*M-H*) behavior of PZT/CFO/PZT tri-layered nanostructure in the first quadrant. (Inset shows the complete loops for selected temperatures). Note the difference in the trend of change in magnetization with temperature before and after 7 kOe.

Figure 3: Magnified *M-H* isotherms in temperature range of 200 to 400 K for applied fields (a) below and (b) above 5 kOe. Inset shows the *M-H* isotherms for all temperatures above 100 K. Star symbol denotes the *M-H* isotherms independent of MEC in respective applied field range.

Figure 4: The variation in entropy *(ΔS(H))* as a function of temperature for fields applied (a) below 0.5 T (b) above 0.5 T (calculated using Maxwell equations). The variation in MCE temperature change (Δ*T)* as a function of temperature for selected applied fields (c) below 0.5 T (d) above 0.5 T. The negative values indicate a positive MCE while the positive values denote the negative/inverse MCE.

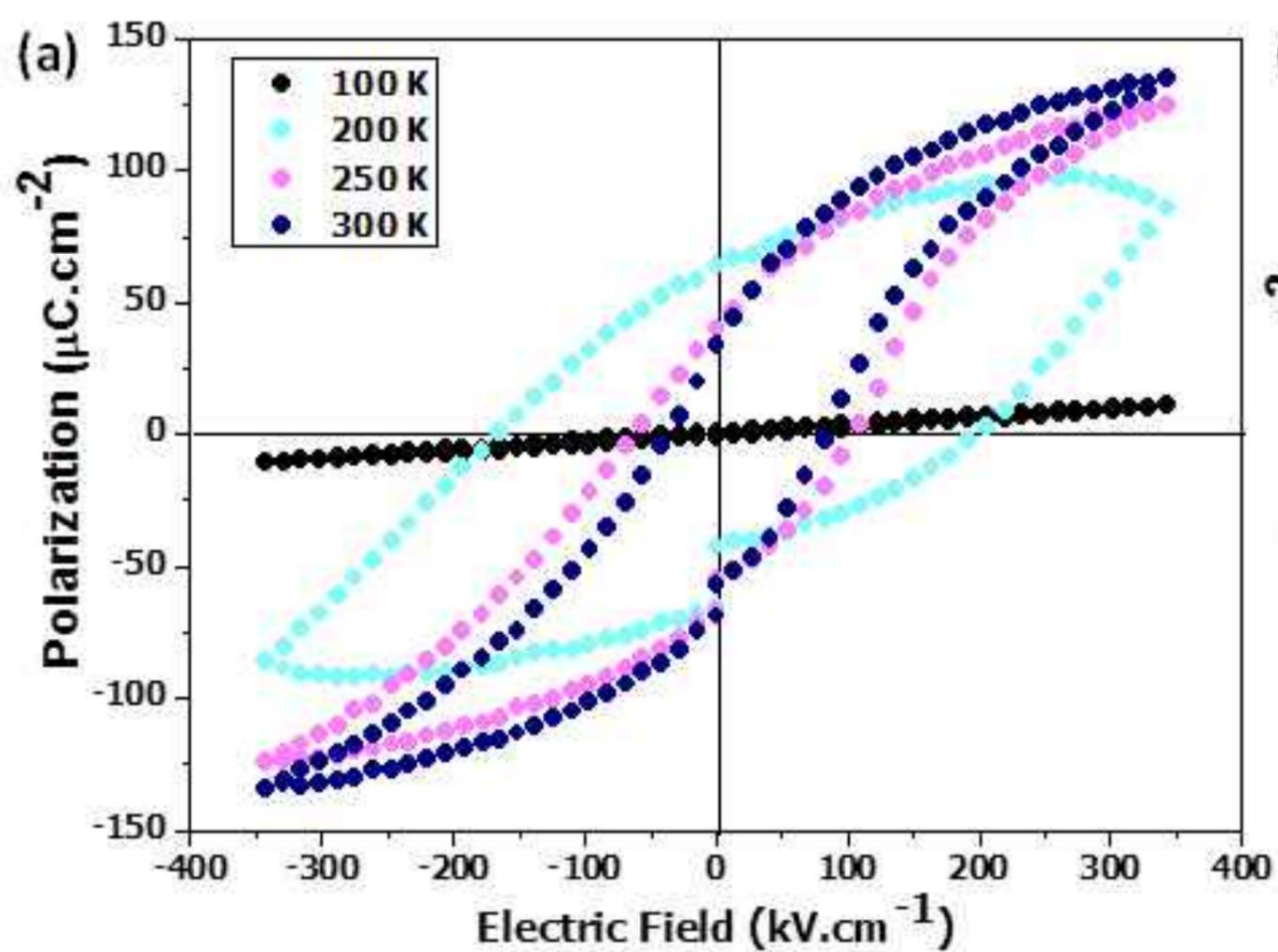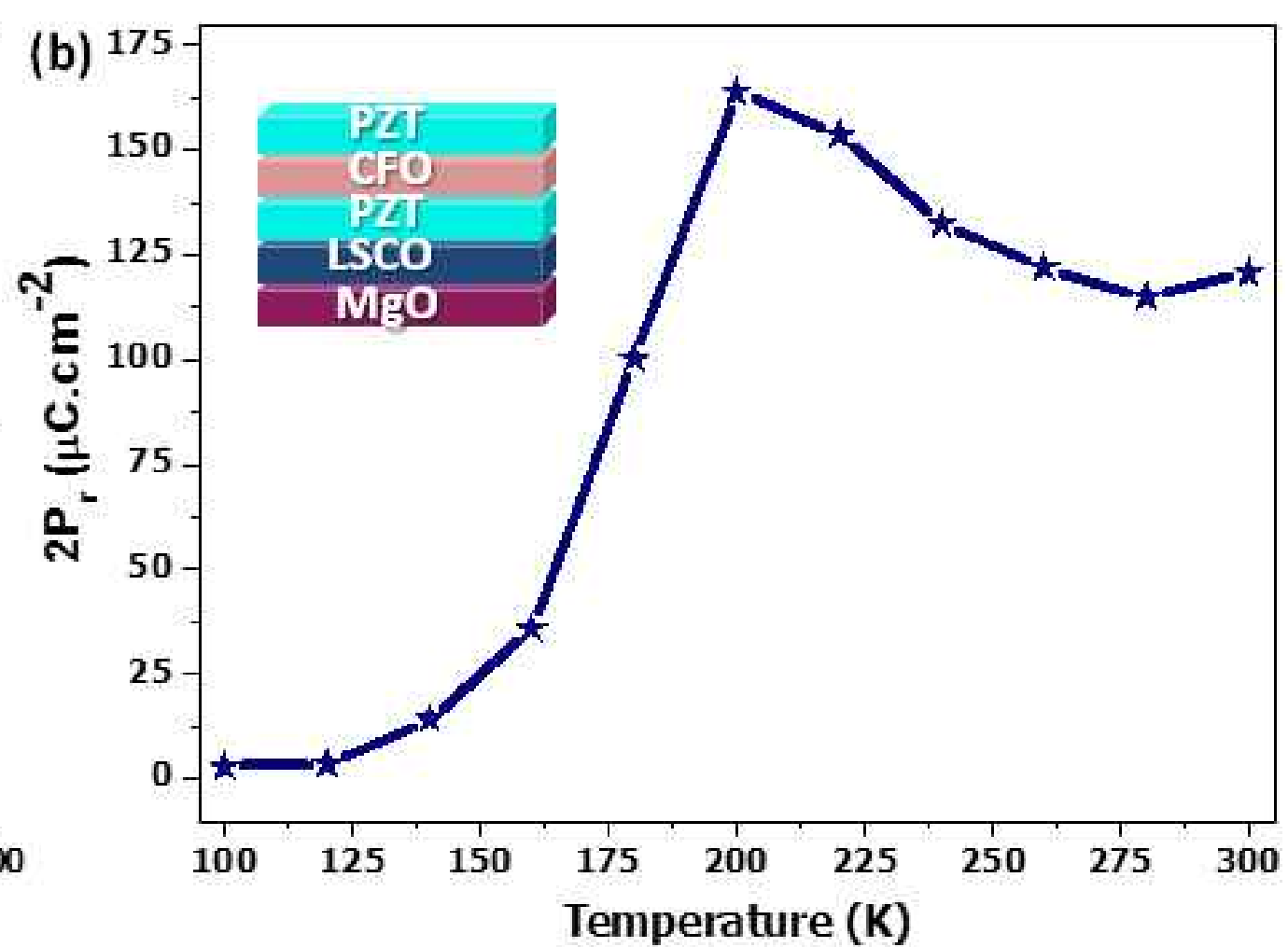

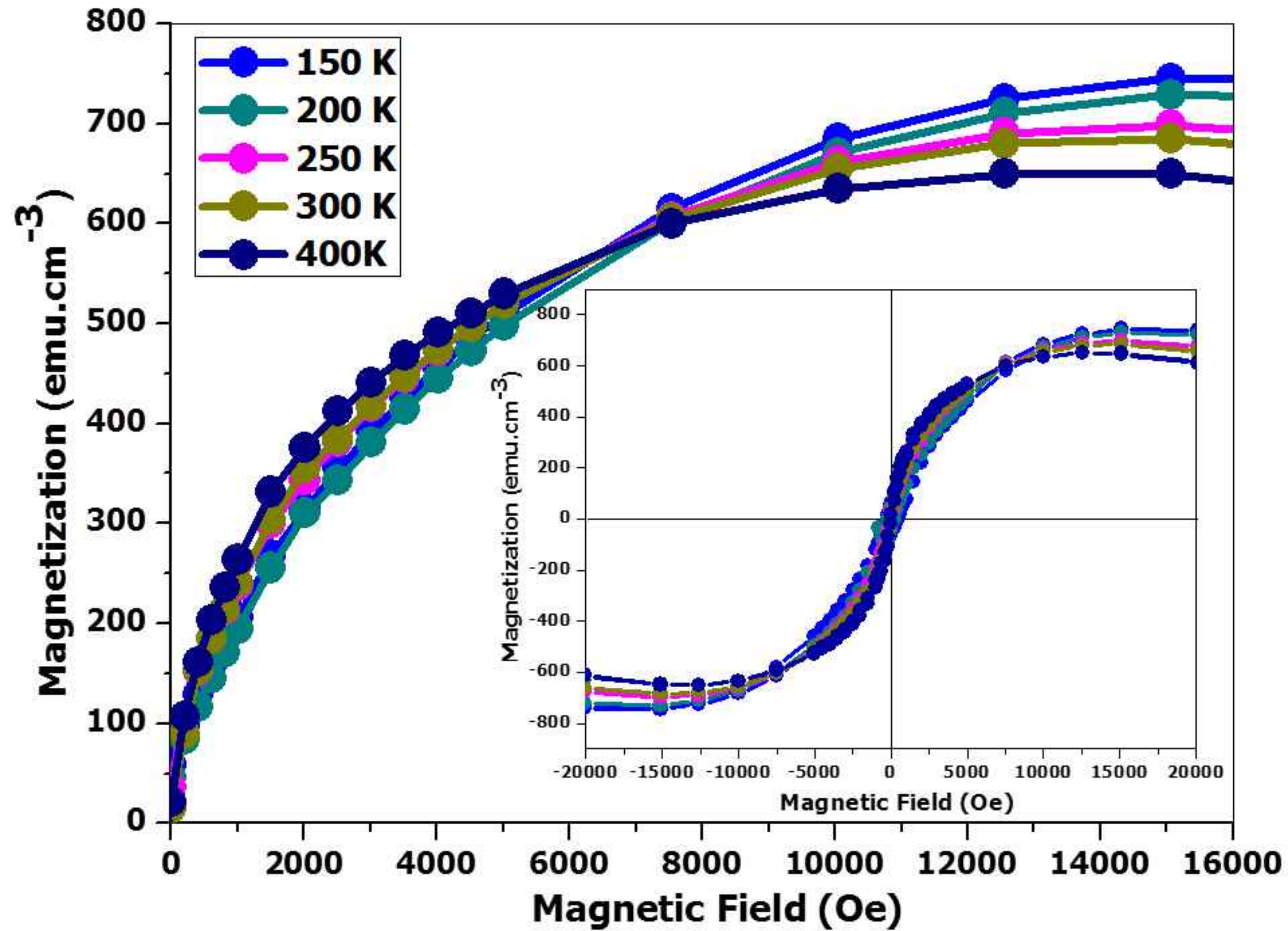

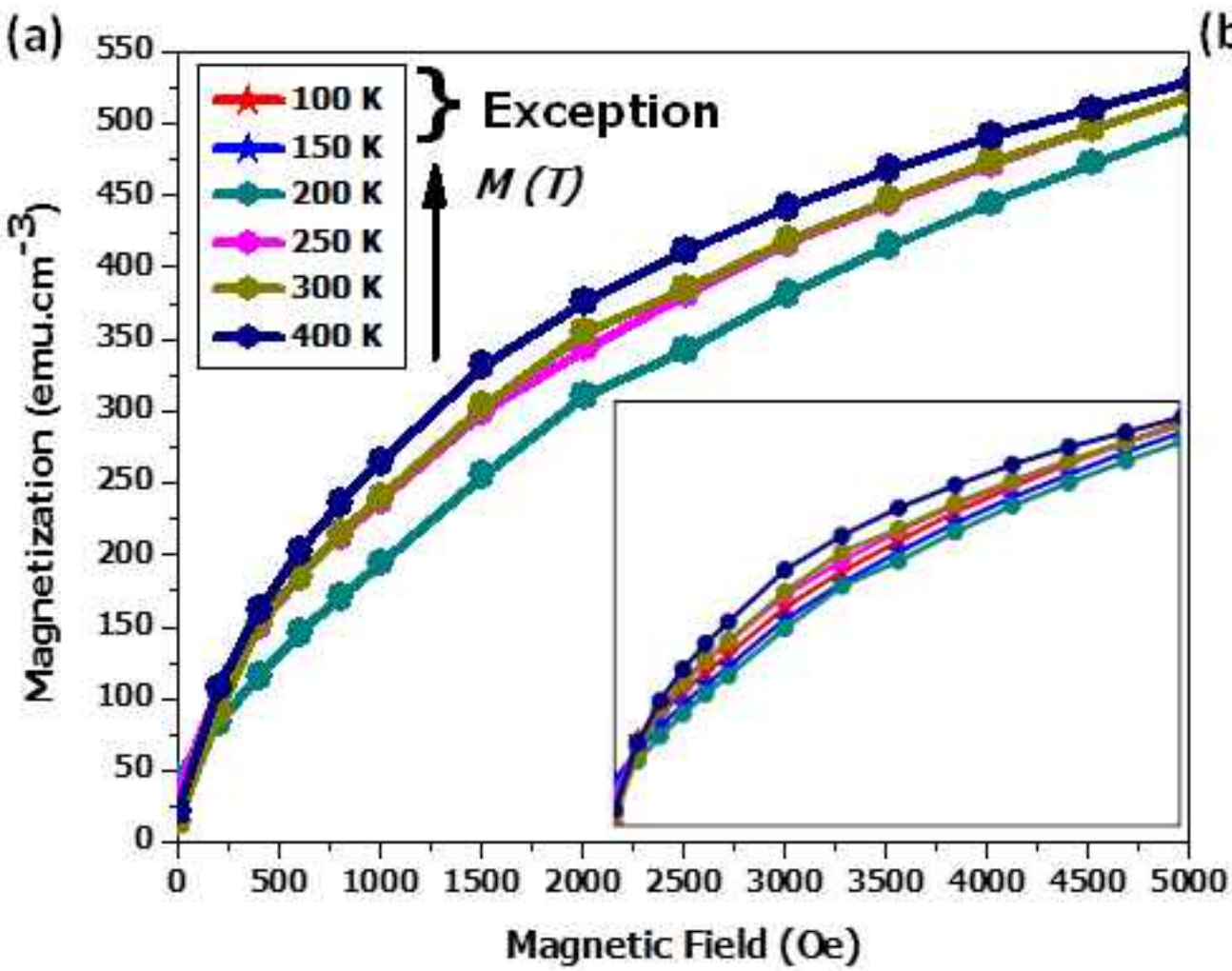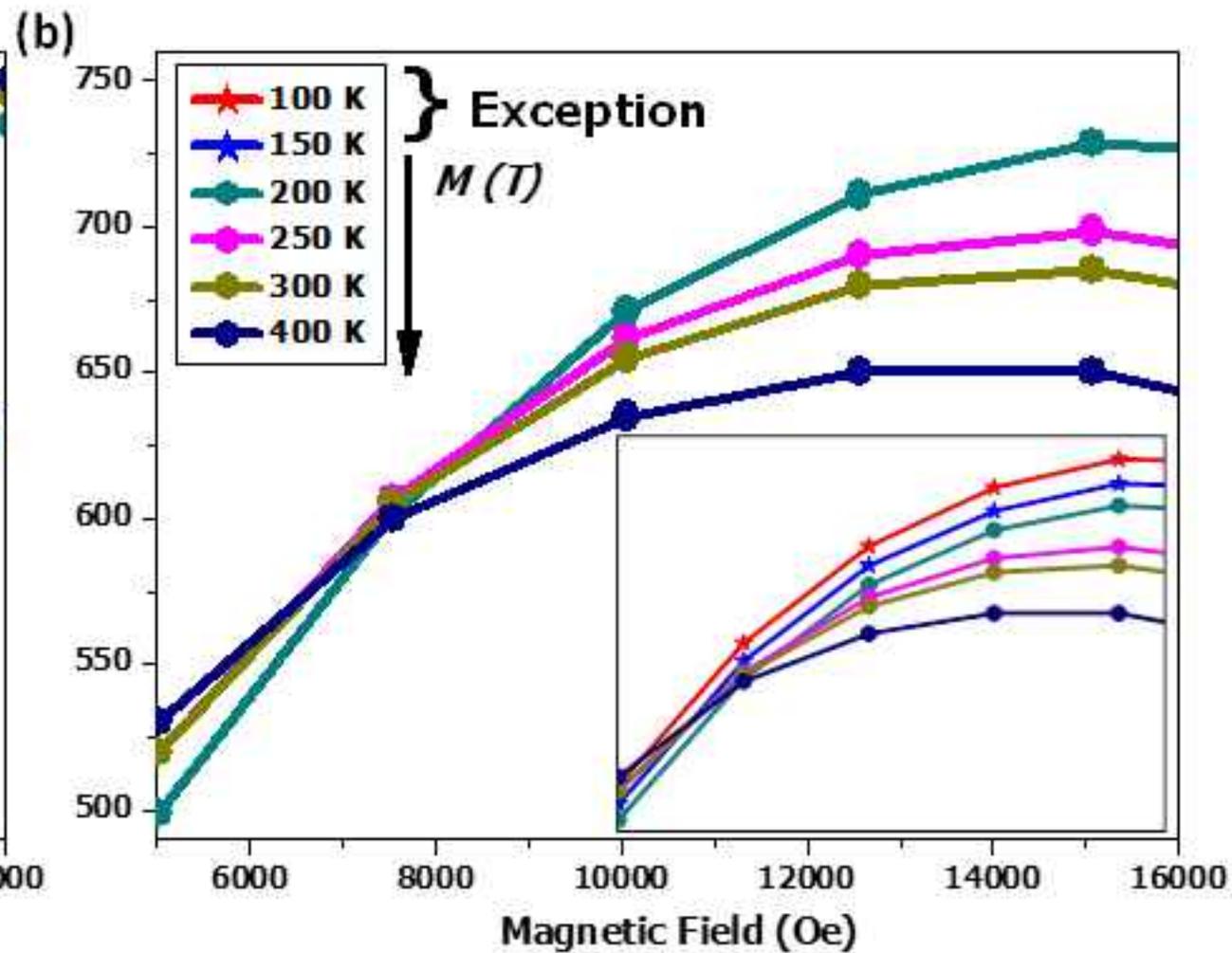

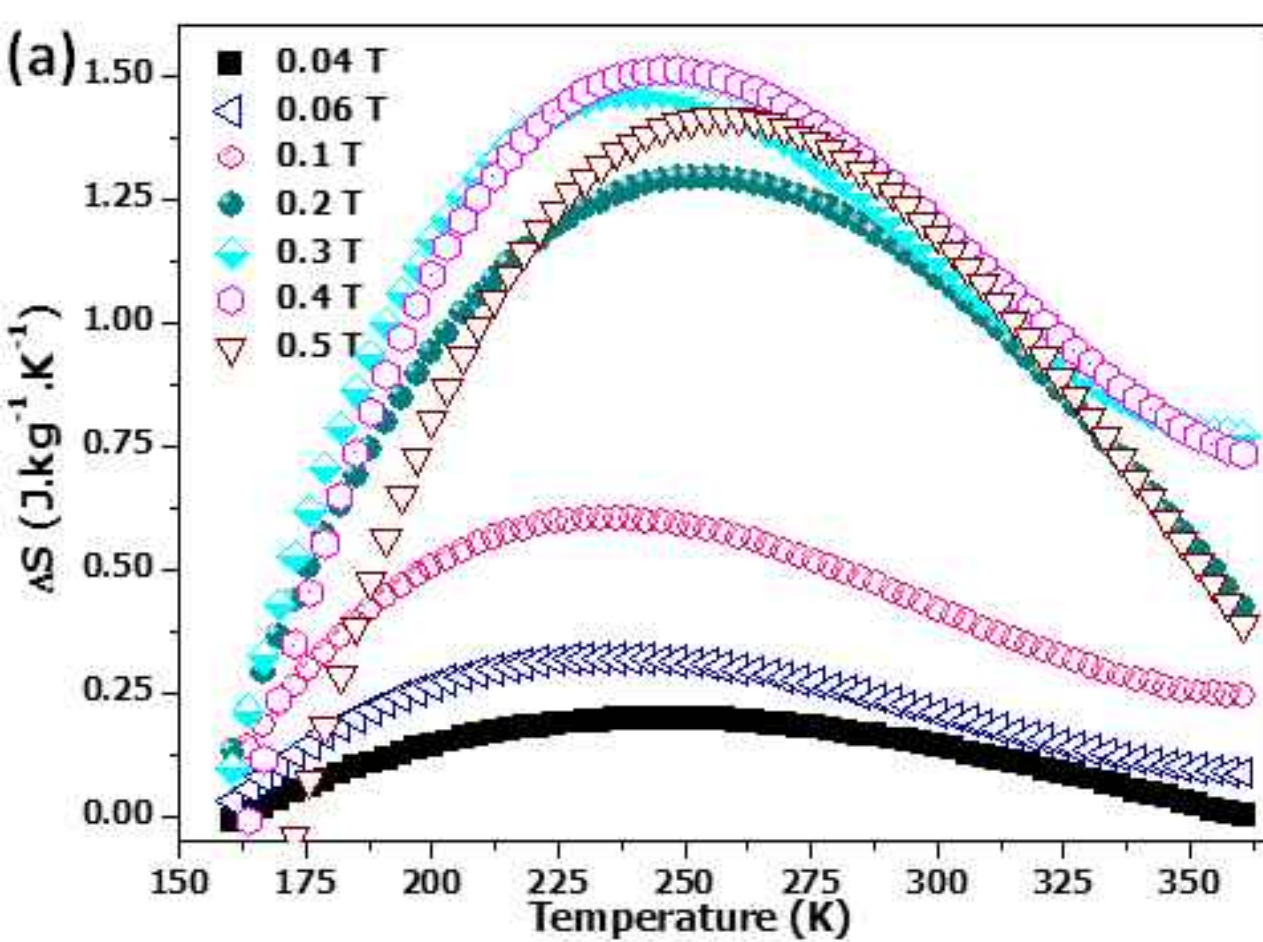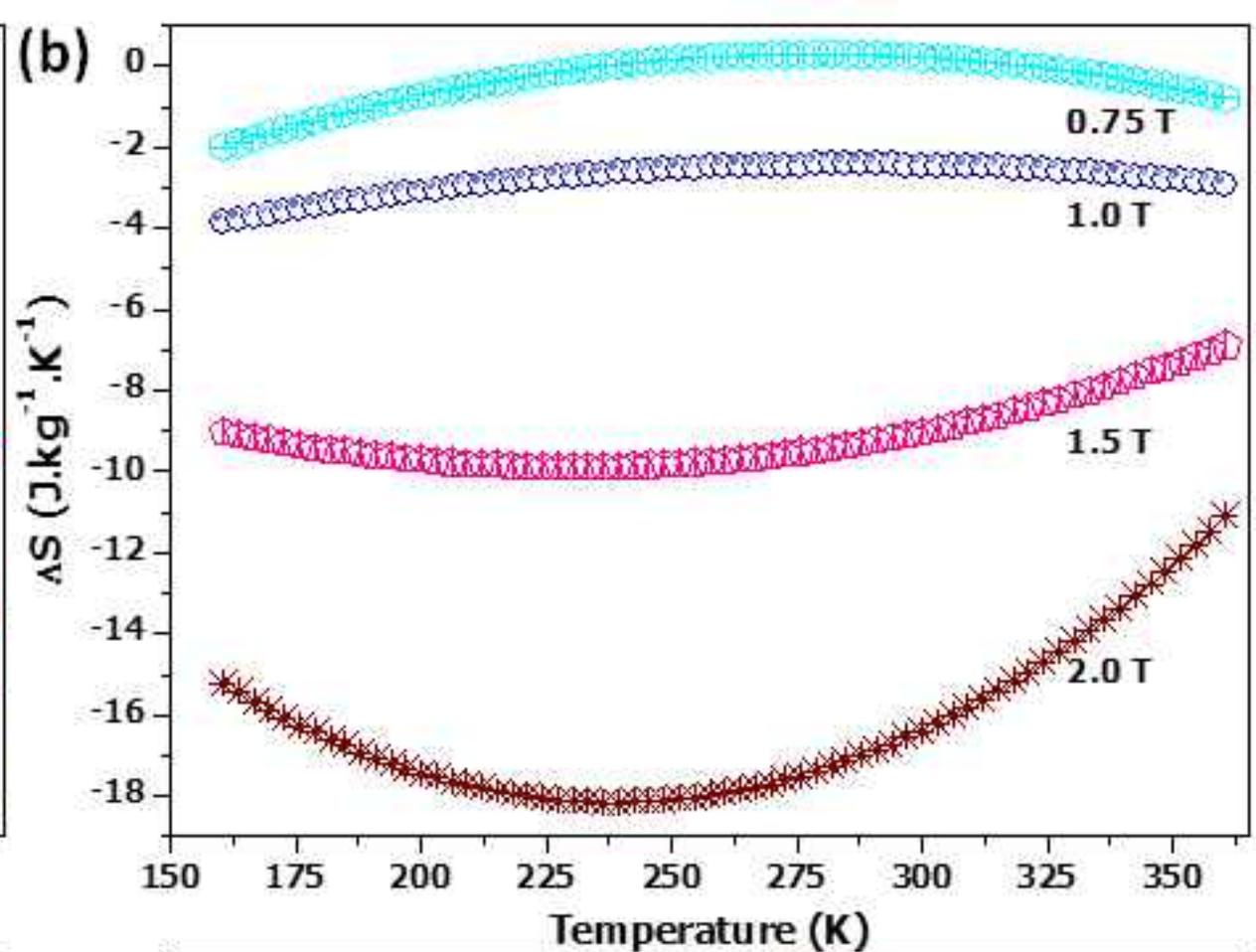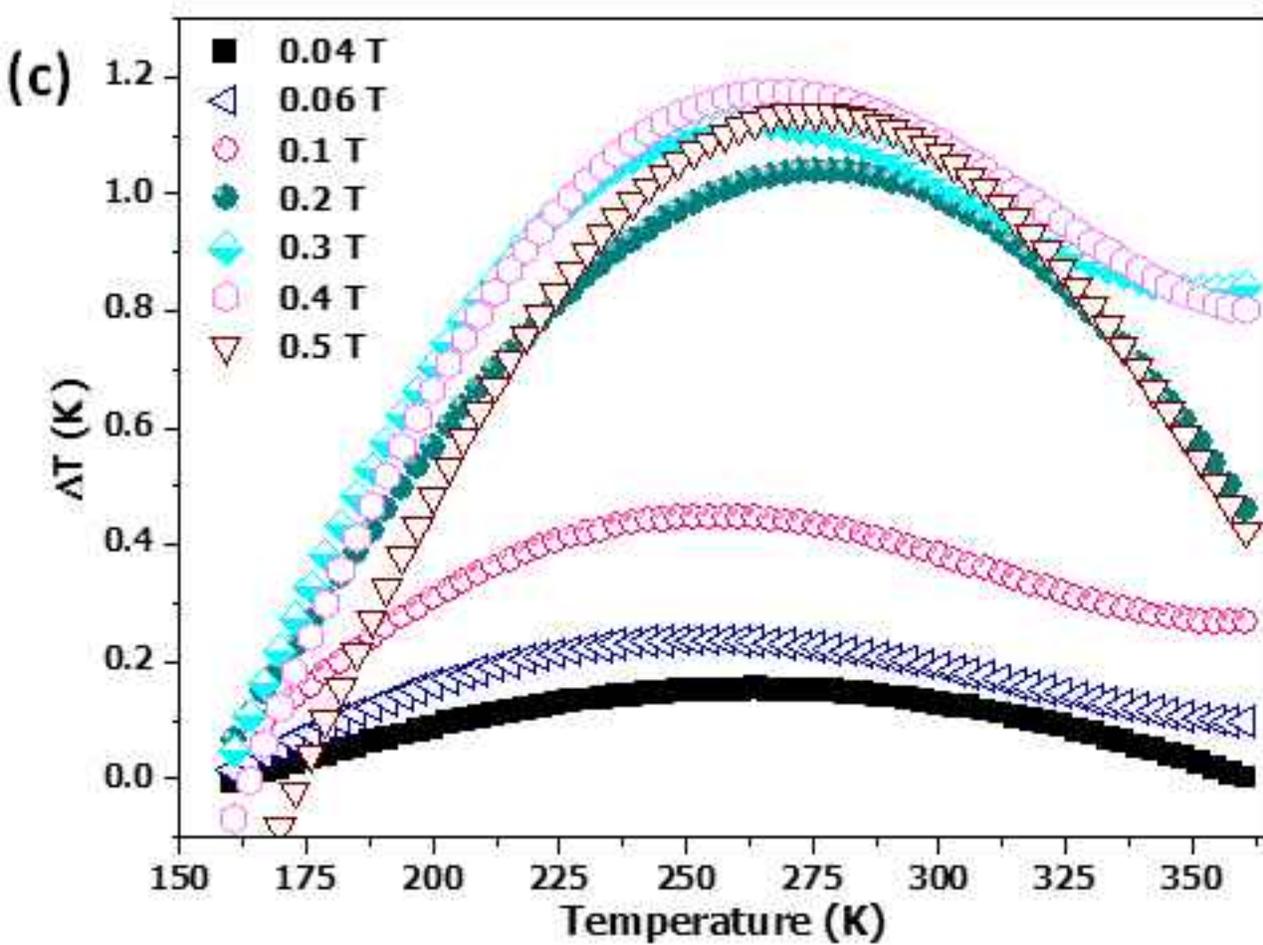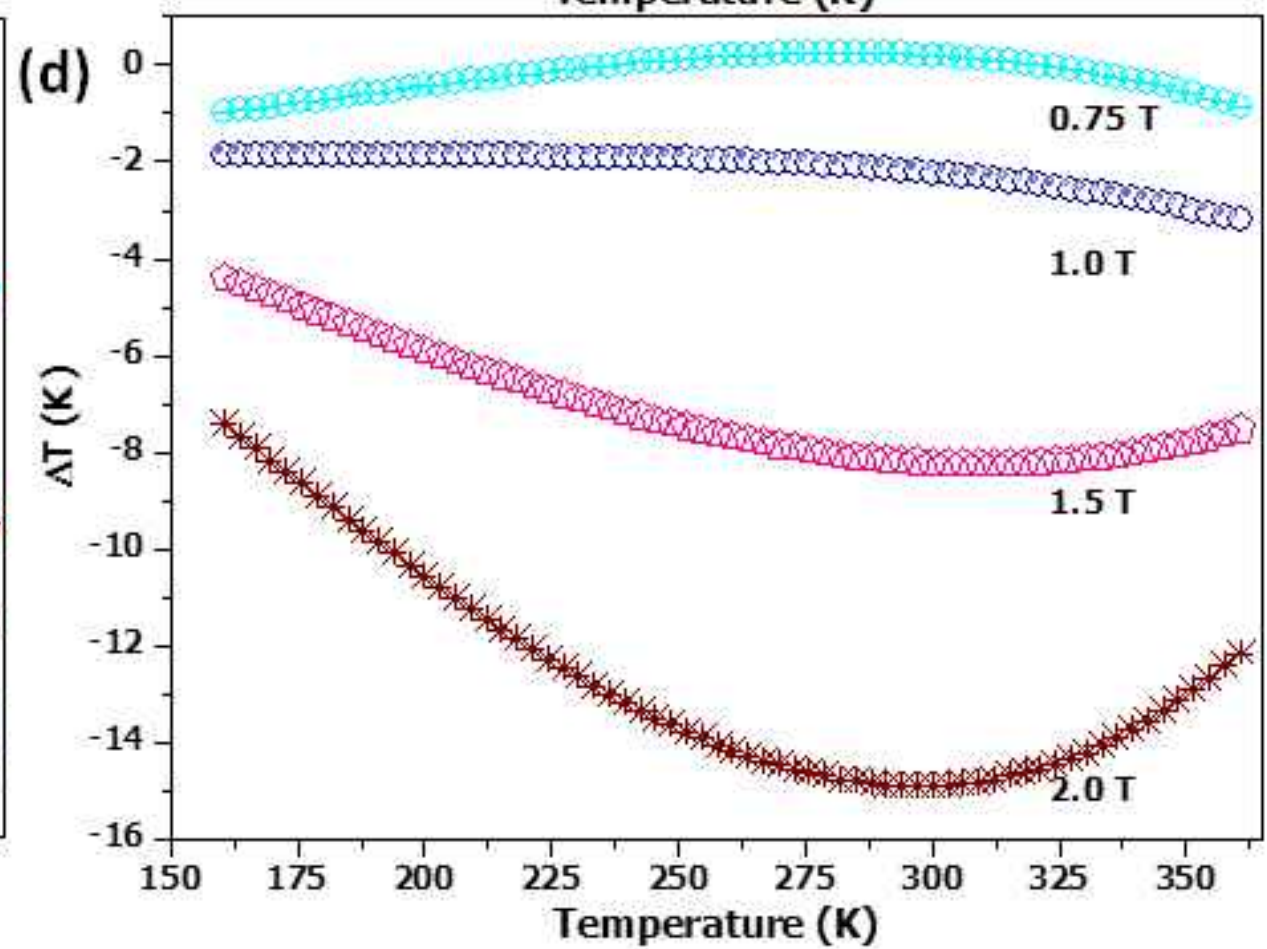